\documentclass{article}

\usepackage{arxiv}
\usepackage[utf8]{inputenc} 
\usepackage[T1]{fontenc}    
\usepackage{epstopdf}
\usepackage{cite}
\usepackage{url}
\usepackage{amsmath,amssymb,amsfonts}
\usepackage{xargs}
\usepackage{graphicx}
\usepackage{paralist}
\usepackage{textcomp}
\usepackage[dvipsnames]{xcolor}
\usepackage[hidelinks,pdftex]{hyperref}
\usepackage[colorinlistoftodos,prependcaption,textsize=tiny]{todonotes}
\usepackage{subfig}
\usepackage{listings}
\usepackage{enumitem}
\usepackage{orcidlink}

\newcommand{\change}[1]{\textcolor[rgb]{0.0,0.0,0.0}{#1}}

\newcommand{\eg}{{\textit{e.g., }}}
\newcommand{\ie}{{\textit{i.e., }}}

\newcommandx{\thought}[2][1=]{\todo[inline,backgroundcolor=white, bordercolor=white, textcolor=black!85,#1]{\small \emph{\{~#2~\} }}}
\newcommandx{\summary}[2][1=]%
    {\vspace{1em} \todo[inline,backgroundcolor=white, bordercolor=black!35, textcolor=black!75,#1]
     {\small \emph{ Summary:~#2 }}}

\newcommand{\marbl}{\textsc{Marbl}}

\title{Comparing Cross-Platform Performance via Node-to-Node Scaling Studies}
\author{Kenneth Weiss$^\dagger$, Thomas M. Stitt$^\dagger$, Daryl Hawkins$^\dagger$,\\
  Olga Pearce$^{\dagger*}$, Stephanie Brink$^\dagger$, Robert N. Rieben$^\dagger$, \\
  \\
  Lawrence Livermore National Laboratory, Livermore, CA, USA$^\dagger$, \\Texas A\&M University, College Station, TX, USA$^*$
}

\begin{document}
\maketitle

\begin{abstract}
    Due to the increasing diversity of high-performance computing architectures, researchers and practitioners are increasingly interested 
    in comparing a code's performance and scalability across different platforms.
    However, there is a lack of available guidance on how to actually set up and analyze such cross-platform studies.
    In this paper, we contend that the natural base unit of computing for such studies is a single compute node on each platform
    and offer guidance in setting up, running, and analyzing node-to-node scaling studies.
    We propose templates for presenting scaling results of these studies and provide several case studies
    highlighting the benefits of this approach.
\end{abstract}


\section{Introduction}

With the culmination of the Exascale Computing Project (ECP)~\cite{messina-2017} 
and the successful deployment of the first exascale machines\change{---most recently,
El Capitan at Lawrence Livermore National Laboratory with 1.7 exaflops---}
there is now considerable diversity in the node architecture of supercomputers.
Of the systems on the \change{November} 2024 Top500 list~\cite{Top500-Nov2024},
\change{29.2\% have both GPUs and CPUs,
totaling an impressive 41.3\% of performance share}.
Innovations in hardware are driving that diversity further,
fueled at least in part by the explosion of machine learning workloads.
%
%
At the same time, simulation code teams are tasked with running performantly on multiple hardware platforms, 
ranging from state-of-the-art CPU-based and GPU-based supercomputers, to commodity clusters and even consumer laptops. 
This is especially challenging when working on large codebases, which must simultaneously support a variety of existing and upcoming architectures, 
as well as new features, and it is infeasible to develop, manage, test, and maintain platform-specific ports of the codebase.
Many teams have risen to this challenge through a single-source
performance-portable port via abstraction libraries such as
\begin{inparaitem}[]
    \item RAJA~\cite{hornung2014raja,beckingsale2019raja},
    \item Kokkos~\cite{edwards2014kokkos,trott-2022},
    \item SYCL~\cite{alpay-2022} and OpenMP, among others~\cite{herten-2023}.
\end{inparaitem}

However, while high-performance computing (HPC) researchers and practitioners now have a plethora of computing platforms to choose from,
it is not always clear which platform to use for solving a desired problem given practical constraints 
involving cluster availability and project budgets.
%
%
Although performance on specific architectures has always been an important consideration in system procurement,
it is especially relevant in the context of cloud computing, where one must select from a range of compute node architectures and
pay for their features accordingly.

Given the increasing importance of this question, there is a surprising lack of guidance in the literature 
for how to actually compare performance across diverse systems.
For example, given the differences in node architectures and features, 
how should one set up scaling studies to compare performance across platforms? 
Similarly, how does one effectively analyze and present the results of such a study?

In this paper, we contend that the relevant unit of computation for such studies is a single compute node on each platform
and that cross-platform studies should be compared at the node level, even when simulations do not fully utilize all the available features of each node.
We discuss how to set up such \emph{node-to-node} scaling studies and propose templates for plotting scaling results to help analyze and compare cross-platform performance.
We also discuss several case studies detailing how charts of this form have
helped understand code performance.
We aim to spark further discussion on the topic, and showcase future studies in follow-up work.

\change{The remainder of this paper is organized as follows:}
Section~\ref{sec:background} provides an overview of scaling studies and current
state-of-the-art. Section~\ref{sec:n2n-scaling-studies} discusses different types of node-to-node
scaling studies. Section~\ref{sec:workflow} details a workflow for running a node-to-node
scaling study. A series of case studies using node-to-node scaling studies to
understand code performance are described in Section~\ref{sec:studies},
\change{followed in Section~\ref{sec:conclusion} with concluding remarks.}

\section{Background and Related Work}
\label{sec:background}


Fundamentally, scaling studies consider how \emph{performance} changes
when running a parameterized \emph{problem} in parallel on one or more \emph{compute resource}.
There are several variables in this statement that need to be unpacked.
\emph{Performance} can include measures of time, energy/power, memory and network efficiency, among other quantities of interest.
The \emph{problem} parameters can include the number of degrees of freedom (``problem size''),
either locally, on each compute resource, or globally for the entire problem.
It can also include aspects of the methodology, for example, when comparing improvements due to different algorithms
or numerical discretizations (\eg finite elements vs. finite volumes) or mesh adaptivity.
For high-order codes it could also include the polynomial order of the mesh and/or solution space.
Similarly, \emph{compute resources} can include node architectures (\eg CPU vs. GPU),
problem distributions across the compute resources (when using MPI or OpenMP)
or different strategies for overloading/underloading the compute resource,
\eg for improved access to shared resources like memory or network.

Scaling studies are typically run to answer questions such as:
\begin{itemize}
    \item How quickly can we turn around a problem?
    \item How large of a problem can we reasonably solve, \eg for convergence studies?
    \item How many Degrees of Freedom (DOFs) should we run on each compute resource?
    \item \change{How does performance compare on different platforms, such as one
with lower latency versus one with higher bandwidth?}
\end{itemize}
Limiting factors for parallel scalability include the portion of the code that
has not been parallelized~\cite{amdahl-67}
as well as communication overhead~\cite{gustafson-88} and the locality of the computations.
\cite{calotoiu-ea:2013:modeling} discuss performance modeling for discovering and resolving scalability bugs.

The traditional baseline for comparisons in a scaling study is a single processor on a compute node.
While this notion has been broadly popularized~\cite{hpc-wiki-scaling, pdc-blog-scaling, cornell-virtual-workshop-scaling},
and used in many strong and weak scaling studies~\cite{eager1989speedup, NICOL1988404, grama1993isoefficiency, singh-93, sun-1990, sun-1991},
it is difficult to compare a code's performance across heterogeneous compute types
at the granularity of a single processor.
We discuss different types of scaling studies and their node-to-node counterparts in Section~\ref{sec:n2n-scaling-studies}
and direct interested readers to \cite{hoefler2015scientific} for an excellent discussion of important desiderata
when discussing and presenting performance results.

For cross-platform studies, we are typically interested in comparing performance
on several compute resources with different architectures.
These studies can take several forms:
\begin{description}
    \item [Node-to-node comparisons]
            In our experience, this is typically how developers track their progress when porting codes
            as it is a pragmatic measure for the impact of code improvements.
            For example, when porting a CPU code to a GPU context, developers often have a speedup metric in mind,
            \eg achieving a 10x speedup. Since users in HPC and cloud computing centers are typically charged per node,
            this is also how HPC users tend to think about their runs.
    \item [Time-to-solution] This measures how quickly problems can be turned around to achieve a desired result.
            Thus, a different way of considering cross-platform performance would be to consider
            the optimal parameters and node counts required to solve a problem on a given platform.
            There are often additional tradeoffs to consider, such as the available resources and computational
            budget, especially when considering the HPC center as a whole or for an ensemble of computations
            where users are interested in optimizing the turnaround times/throughput for a collection of computing jobs.
    \item [Energy/power usage] Energy concerns are an increasingly relevant concern for HPC users.
            In this case, one might consider the power/energy cost of running the simulation, even if
            this increases the time-to-solution.
            Parallel architectures, such as those with GPUs, use more (instantaneous) power, but
            their associated speedups can turn around a job faster, thereby using less overall energy
            (integral of power over time)~\cite{Ryujin2022-arxiv, horwitz2024estimating, patki-2021}.
    \item [Code-to-code comparisons] If we have more than one code that can solve a given problem, we might
            be concerned with which code to run when solving a given task on a given platform. A different spin
            on this question is intra-code comparisons using different methodologies or discretization parameters.
\end{description}
%

Cross-platform scaling studies require careful coordination and bookkeeping. Several workflow management tools and scheduling managers have been developed in recent years to help manage ensembles of simulations%
~\cite{maestro, DiNatale-2019-ams,peterson2019merlin,jacobsen2023ramble,ahn-2020-flux},
however, they do not all have built-in support for running simulations on different clusters and for comparing the results.


Similarly, there are many tools to analyze a code's performance, often focusing on a single architecture,
\eg CPU, GPU, or a single aspect of performance such as MPI, OpenMP, CUDA or HIP.
Examples include
\begin{inparaitem}[]
    \item PAPI counters~\cite{papi}, a long-standing effort to standardize measurements on CPUs;
    \item ARM forge~\cite{armforge};
    \item Intel V/Tune and Top-down analysis~\cite{yasin2014top};
    \item NVIDIA Nsight~\cite{ncu} and roofline analysis~\cite{gpu-rooflines}; and
    \item platform-independent tools like Totalview,
HPCToolkit~\cite{adhianto2010hpctoolkit} and TAU~\cite{10.1177/1094342006064482}.
\end{inparaitem}
In contrast, Benchpark~\cite{pearce-23-benchpark} is a recent tool for benchmarking application performance, 
on different platforms, especially in the context of system acquisitions.

The recently introduced \emph{Ubiquitous Performance Analysis} paradigm~\cite{Boehme2021-isc} exposes simulation codes
to multiple performance measuring services.
In this paradigm, a code is instrumented with explicit semantically meaningful
annotated regions using Caliper~\cite{boehme:sc2016} and metadata using
Adiak~\cite{adiak}, and runtime-enabled services control what is measured and exported.
The output can then be analyzed using script-based post-processing tools, such
as Hatchet~\cite{bhatele2019hatchet} and
Thicket~\cite{brink-2023-thicket}.
While Hatchet focuses on comparing pairs of runs, Thicket can compare ensembles of runs.
Since performance measurements are built into the code and always available, this paradigm reduces the friction for developers to measure performance.

%

\section{Node-to-Node Scaling Studies}
\label{sec:n2n-scaling-studies}

In this section, we discuss aspects related to designing several types of node-to-node scaling studies
and provide mockups for charts to aid in their analysis.

\subsection{Node-to-Node Strong Scaling Studies}

    In a strong scaling study, we keep the total problem size fixed 
    and vary the computational resources used to solve the problem.
    For node-to-node strong scaling studies, each run often doubles the number of compute nodes in the previous run.
    Thus, successive runs in the series solve fewer degrees of freedom on each compute node.

    Considering a single platform $P$, we are mainly interested in the \emph{strong scaling speedup}
    obtained when running on $N$ nodes of $P$ compared to running on a single node on the same platform.
    That is, if it takes $t_{P}(1)$ time to run the problem on a single compute node of $P$
    and $t_{P}(N)$ time to run on $N$ nodes of $P$,
    then the strong scaling speedup for the $N$ nodes is $t_{P}(1)/t_{P}(N)$.
    In other words, the runtime for ideal strong scaling decreases linearly in the number of nodes.

    The limiting factors for node-to-node strong scaling include
        serial effects as in traditional strong scaling~\cite{amdahl-67}:
        as more code is parallelized, the serial portion of the code accounts for relatively larger percentages of the runtime.
    Strong scaling speedups for simulation codes can also be affected by increased communication overhead over the simulation's
    boundary elements since, at higher node counts, each rank has fewer elements and thus, more of its elements are on domain boundaries.
    CPU-based platforms are typically able to achieve better strong scaling speedups at higher node counts than
    their GPU-based counterparts since, for a fixed problem size, there is insufficient work to saturate the GPU.
    For GPU-based architectures, factors like kernel launch times and memory movement between host and device
    memory spaces can also \change{significantly} affect speedup factors.

    \begin{figure}[tbp]
    \centering
        \includegraphics[width=0.6\columnwidth]{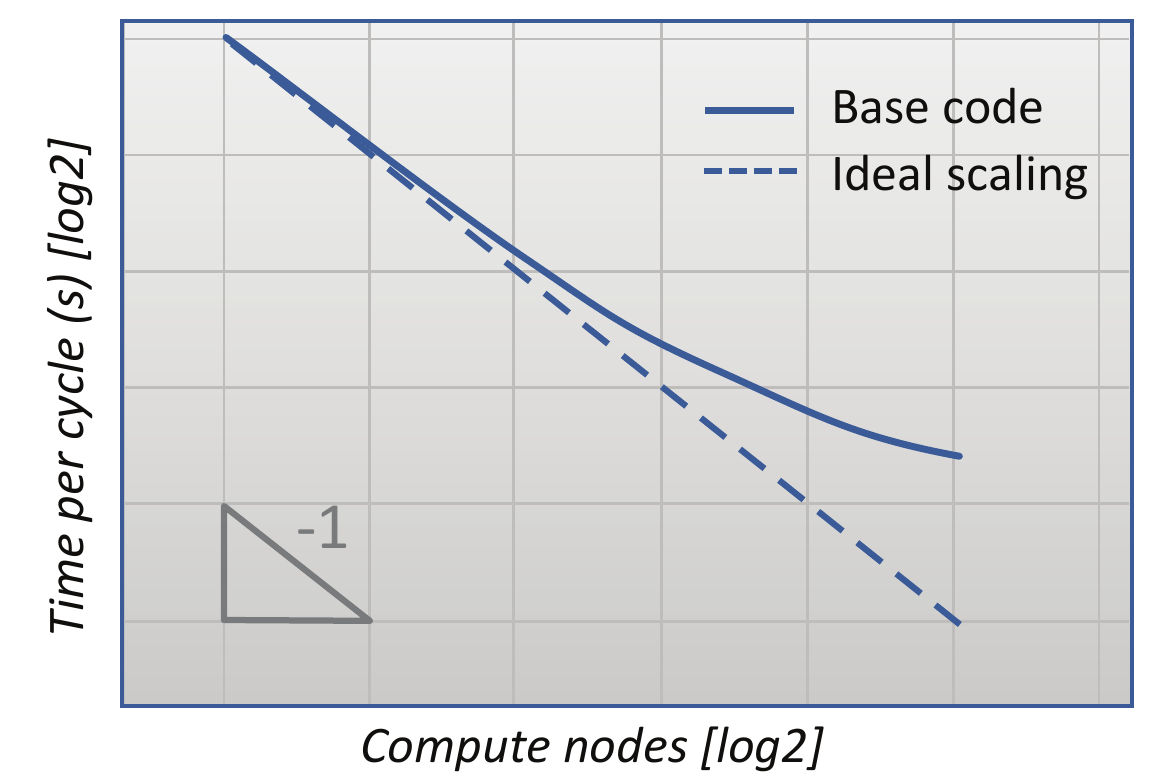}
        \caption{Proposed chart for node-to-node strong scaling studies plotting time per cycle (in seconds)
            against the number of compute nodes. Under ideal strong scaling,
            runtime decreases proportionally with the number of compute nodes,
            with a slope of $-1$ on the $log_2$--$log_2$ axes.}
        \label{fig:strong-scaling}
    \end{figure}

    Figure~\ref{fig:strong-scaling} shows a mockup of our proposed plotting for node-to-node strong scaling studies.
    Salient features include logarithmic ($log_2$--$log_2$) axes,
    which reinforce the notion that as we double the number of nodes,
    an ideally strong-scaled run would complete in half the time.
    For the independent axis, we plot the number of compute nodes
    and for the dependent axis, we plot the time per cycle (in seconds)\change{.}
    Thus, ideal strong scaling \change{(for the general case)} is a diagonal line with slope $-1$.
\change{Superlinear speedup~\cite{556383} is also possible, but is not commonly
achieved.}
    
    By plotting results from multiple platforms, this chart can also be used for cross-platform analysis of strong scaling performance.
    In this case, we are typically interested in the relative performance on each platform when using the same number of nodes.
    For example, the relative speedup when using $N$ nodes of a GPU-based platform $P_{GPU}$
    compared to that of a CPU-based platform $P_{CPU}$ would be $t_{P_{CPU}}(N)/t_{P_{GPU}}(N)$,
    which would correspond to vertical lines in our node-to-node strong scaling chart.
    Alternatively, given a data point on one platform using $N$ nodes,
    we might be interested in seeing the number of nodes $M$ required to achieve similar performance on other platforms,
    \ie a horizontal line in our chart. Other interesting features might be found at crossover points on the chart,
    where two platforms achieve the same performance using the same number of nodes.

\subsection{Node-to-Node Weak Scaling Studies}

    \begin{figure}[tbp]
    \centering
        \includegraphics[width=0.6\columnwidth]{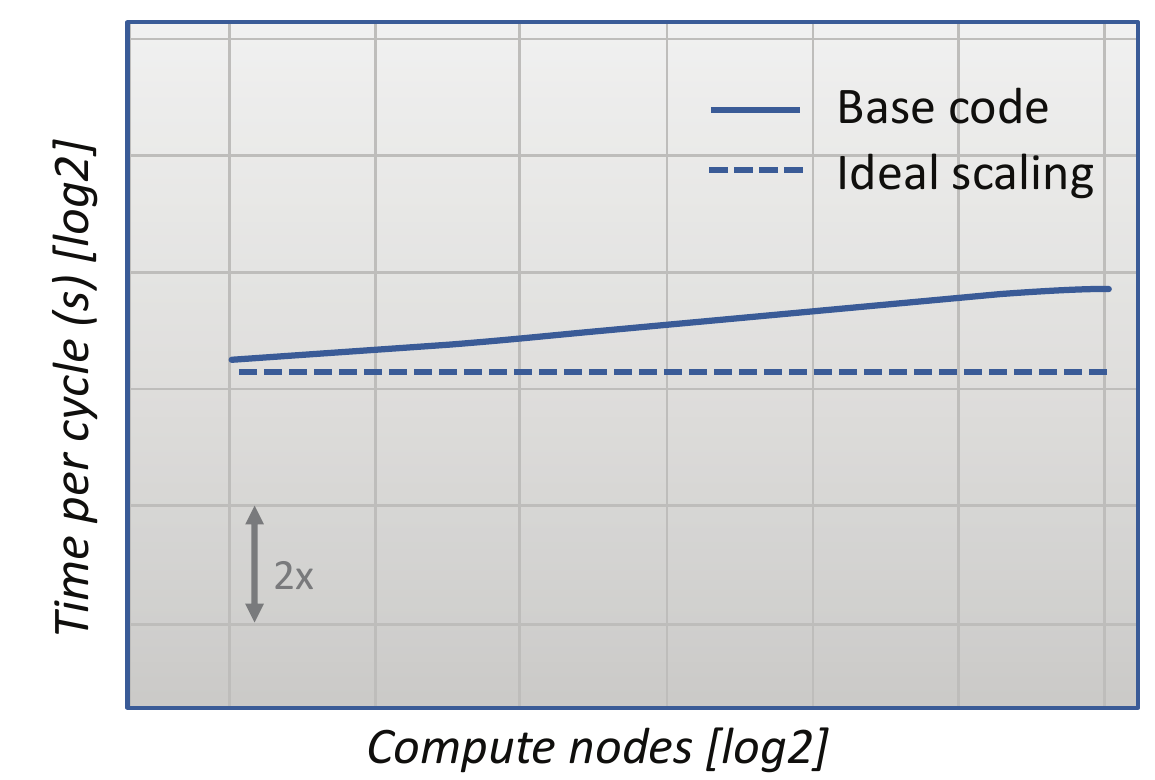}
        \caption{Proposed chart for node-to-node weak scaling studies plotting time per cycle (in seconds) against the number of compute nodes.
            Under ideal weak scaling, runtime is independent of the number of compute nodes, yielding a horizontal line of slope $0$.
            Each horizontal bar corresponds to a factor of two increase in runtime.}
        \label{fig:weak-scaling}
    \end{figure}

    In contrast, a weak scaling study keeps the local problem size fixed on each compute node 
    and varies the total problem size as the number of compute nodes increases. 
    Thus, the amount of work on each compute node remains the same,
    and we can measure factors such as the load imbalance or differences in
network architectures~\cite{gustafson-88}.

    Considering a single platform, $P$, a node-to-node weak scaling study is mainly
    interested in the \emph{weak scaling efficiency}.
    When comparing a run with $N$ nodes that takes $t_P(N)$ time with that of a single node taking $t_P(1)$ time,
    the weak scaling efficiency is $t_P(1)/t_P(N)$.
    Since the per-node work is roughly the same, both runs should ideally take the same amount of time,
    yielding a weak scaling efficiency of $1$.
    This is impractical in practice for MPI-based simulation codes which require intra-process communication
    and typically have some forms of load imbalance.
    The inverse of the weak scaling efficiency can also be used as a way to indicate the \emph{slowdown} experienced as the problem is weak scaled.

    A simple way to set up weak scaling studies on a simulation mesh is to use incremental uniform refinement on its elements.
    That is, we uniformly refine each element of the problem and use a corresponding multiplicative factor on the number of nodes.
    For a 2D problem, we use successive refinements where each element is replaced by four higher resolution elements;
    for 3D, the multiplicative factor is eight.
    This type of weak scaling study grows in node count very rapidly, so one can only obtain a few data points on even large clusters.

    Figure~\ref{fig:weak-scaling} shows our proposed plotting for node-to-node weak scaling studies.
    This chart also uses $log_2$--$log_2$ axes plotting the number of compute nodes against the runtime per cycle (in seconds).
    As such, ideal weak scaling would be a horizontal line of slope $0$.
    \change{The $log_2$ scaling on the dependent axis makes it relatively easy to gauge the performance
    at scale since successive vertical markers indicate a multiplicative speedup factor of two.}

    This chart can also help with cross-platform performance analysis.
    In that context,  we are typically interested in the relative performance among platforms
    for runs that use the same number of nodes.
    Due to the $log_2$ scaling of the dependent axis, this can be easily estimated by counting the number of lines between
    weak scaling curves along vertical lines (\ie fixed number of compute nodes),
    each corresponding to a speedup factor of two.

\subsection{Node-to-Node Strong-Weak Studies}

    \begin{figure}[tbp]
    \centering
        \includegraphics[width=0.6\columnwidth]{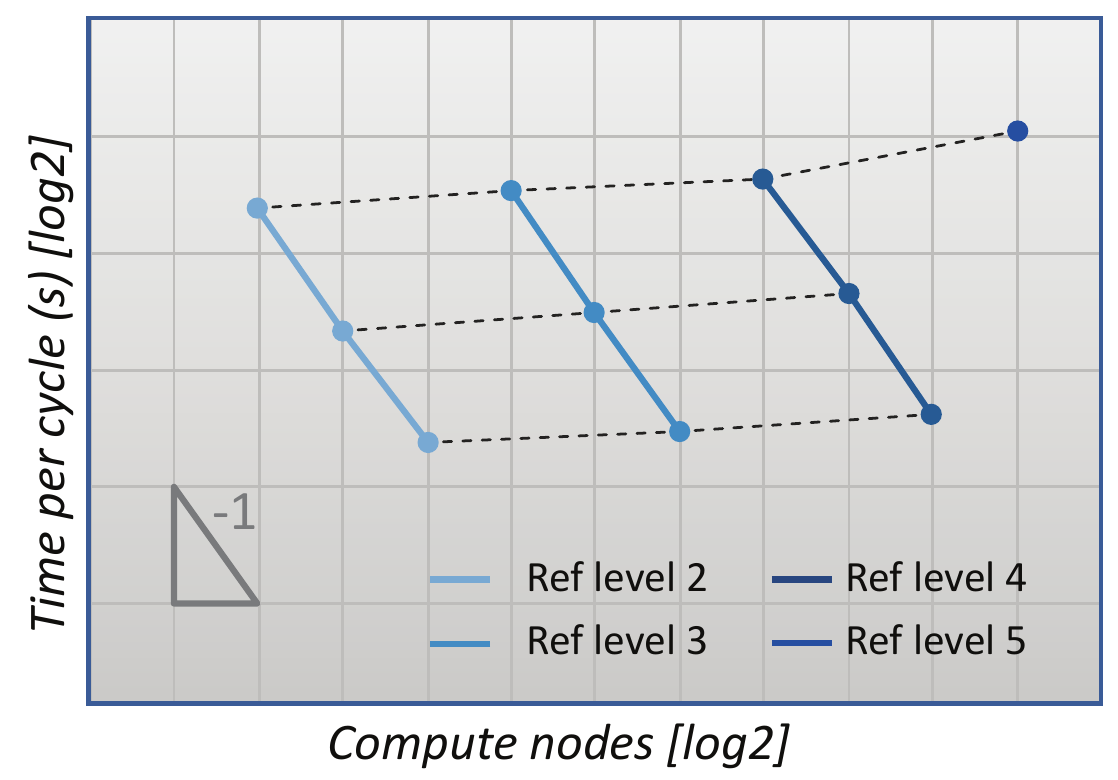}
        \caption{Proposed chart for node-to-node strong-weak studies plotting time per cycle (in seconds) against the number of compute nodes.
                By carefully controlling the study parameters, we can include both strong scaling and weak scaling on the same chart.
                Solid diagonal lines correspond to strong scaling. Dashed horizontal lines correspond to weak scaling.}
        \label{fig:strong-weak-scaling}
    \end{figure}

    By setting up a node-to-node scaling study in the right way, one can plot strong and weak scaling on a single chart.
    Such charts can be useful for figuring out the ``sweet spot'' for running a calculation.
    As an example, if you have two variables, such as number of nodes
    and resolution, to play with, the \emph{strong-weak} scaling chart can give a sense of what to
    expect if you double your compute resource
    or increase/decrease the mesh refinement.
    
    Figure~\ref{fig:strong-weak-scaling} shows our proposed plotting for node-to-node strong-weak scaling studies.
    It uses the same $log_2$--$log_2$ axes as the strong and weak scaling studies,
    but connects strong scaling sample points --- \ie those with the same total
    number of degrees of freedom --- using solid lines,
    and weak scaling data points --- \ie those with (approximately) the same
    number of degrees of freedom per node --- using dashed lines.
    Samples within the same strong scaling series have the same color.
    
\subsection{Node-to-Node Throughput Scaling Studies}
\label{sec:n2n-throughput-scaling-studies}

    \begin{figure}[tbp]
    \centering
        \includegraphics[width=0.6\columnwidth]{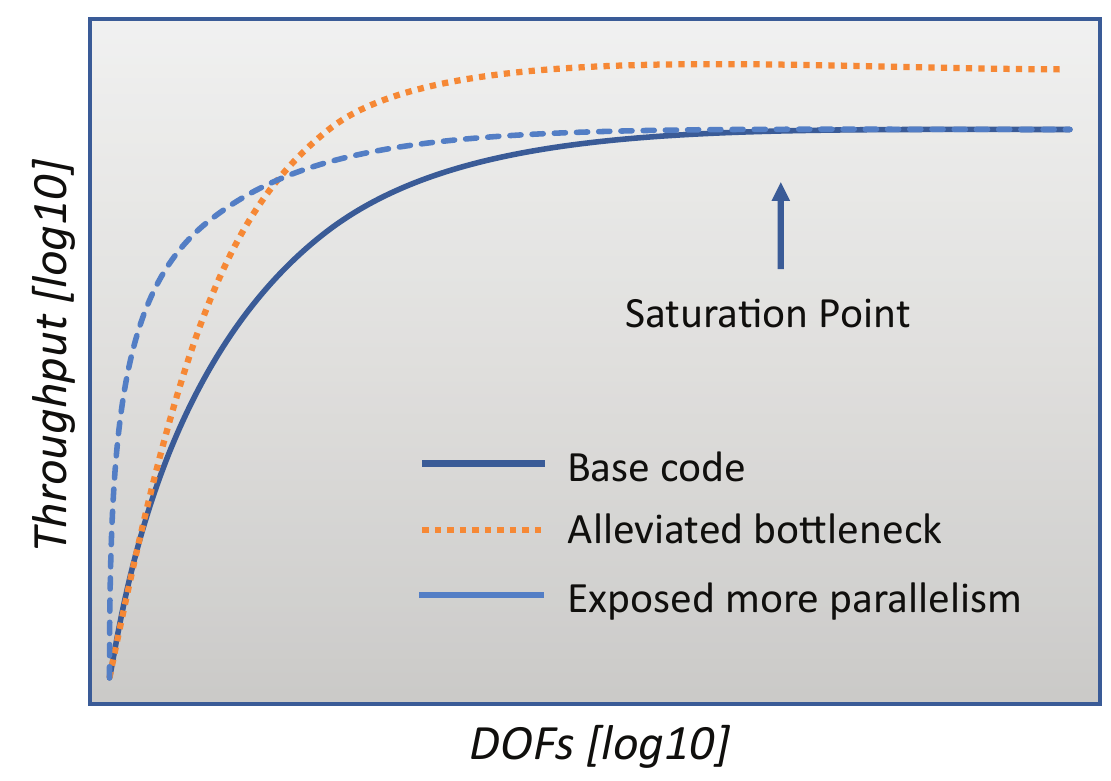}
        \caption{Proposed chart for node-to-node throughput scaling studies
plotting throughput against the problem size (that is, degrees of freedom).
                Throughput helps us understand how well a code is using its compute resources.}
        \label{fig:throughput-scaling}
    \end{figure}

    Throughput studies have a different flavor than strong and weak scaling.
    In this case, we are typically interested in comparing the per-node throughput on a fixed resource, 
    preferably, a single compute node.
    We then vary the number of degrees of freedom in the problem and increase the problem size until we run out of memory (OOM).
    The key metric for these studies is throughput, which is defined as the amount of work done on 
    our compute resource in a given measure of time.
    As an example, we often use the following definition:
    \begin{equation*}
        throughput = \frac{\langle\text{DOFs-processed}\rangle}{\text{compute\_node}} \cdot \frac{cycles}{second}.
    \end{equation*}

    The expectation is that throughput increases with the problem size
    until it hits a \emph{saturation point} for the compute resource, at which point, the resource is bottlenecked.
    Once our resources are saturated, doubling the problem size doubles the time-to-solution.
    There are two strategies to increase performance after saturation:
        (1) expose more parallelism, which could allow the node to saturate
sooner (\ie using a smaller problem size),
        or (2) alleviate bottlenecks in the code, which might raise the saturation point to a higher throughput.
    
    Figure~\ref{fig:throughput-scaling} shows our proposed plotting for node-to-node throughput scaling studies, 
    along with the affect of exposing additional parallelism (blue dashed line) 
    and or alleviating bottlenecks (orange dashed line).
    Similarly to our other node-to-nodes scaling plots, we plot node-to-node throughput with logarithmically-scaled axes.
    However, since throughput does not have a natural connection to powers of two, we use $log_{10}$ scaling instead.    

\subsection{Discussion}

    Strong scaling demonstrates the ability of a code to solve a fixed problem faster as one adds additional compute resources, 
    while weak scaling demonstrates the ability of a code to solve a larger problem faster as one adds more resources. 
    In contrast, throughput scaling characterizes the machine performance and demonstrates the balance 
    between parallelism and memory capacity. It measures the work required to saturate memory bandwidth 
    on the compute resource. 
    Since commodity systems have modest memory bandwidth and compute resources,
    they can only accommodate a relatively small amount of parallelism and
saturate quickly, yielding overall lower performance. 
    On the other hand, GPU-based systems like the Sierra \change{and El Capitan clusters} at Lawrence
    Livermore National Laboratory (LLNL), have very high memory bandwidth and compute resources.
    As such, they require large amounts of parallel work to saturate, but can yield very high performance.

\section{Workflow for Running Node-to-Node Scaling Studies}
\label{sec:workflow}

    \begin{figure}[tbp]
    \centering
            \includegraphics[width=\columnwidth]{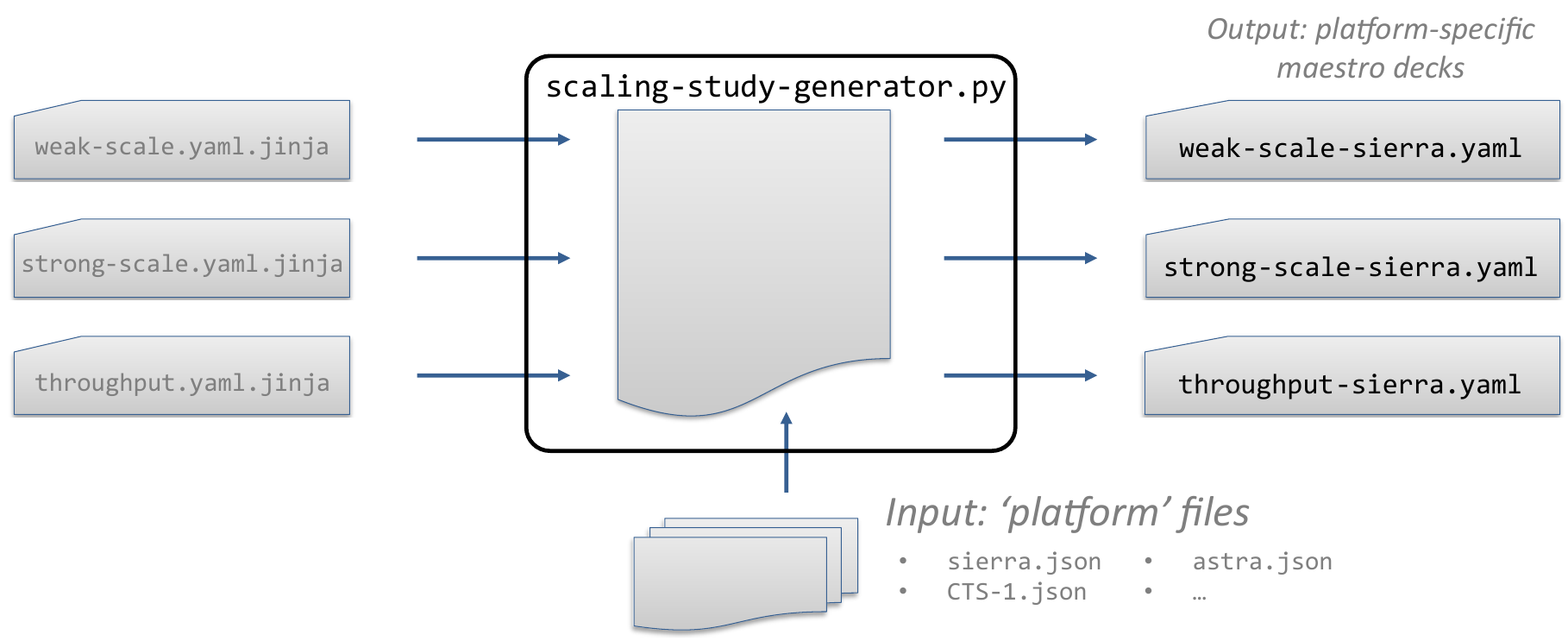}
            \caption{We set up our scaling study parameters and machine details
                    in templated file that are processed into YAML files runnable by Maestro.}
            \label{fig:scaling_study_jinja}
    \end{figure}

    \begin{figure*}[tbp]
    \centering
        \includegraphics[width=\columnwidth]{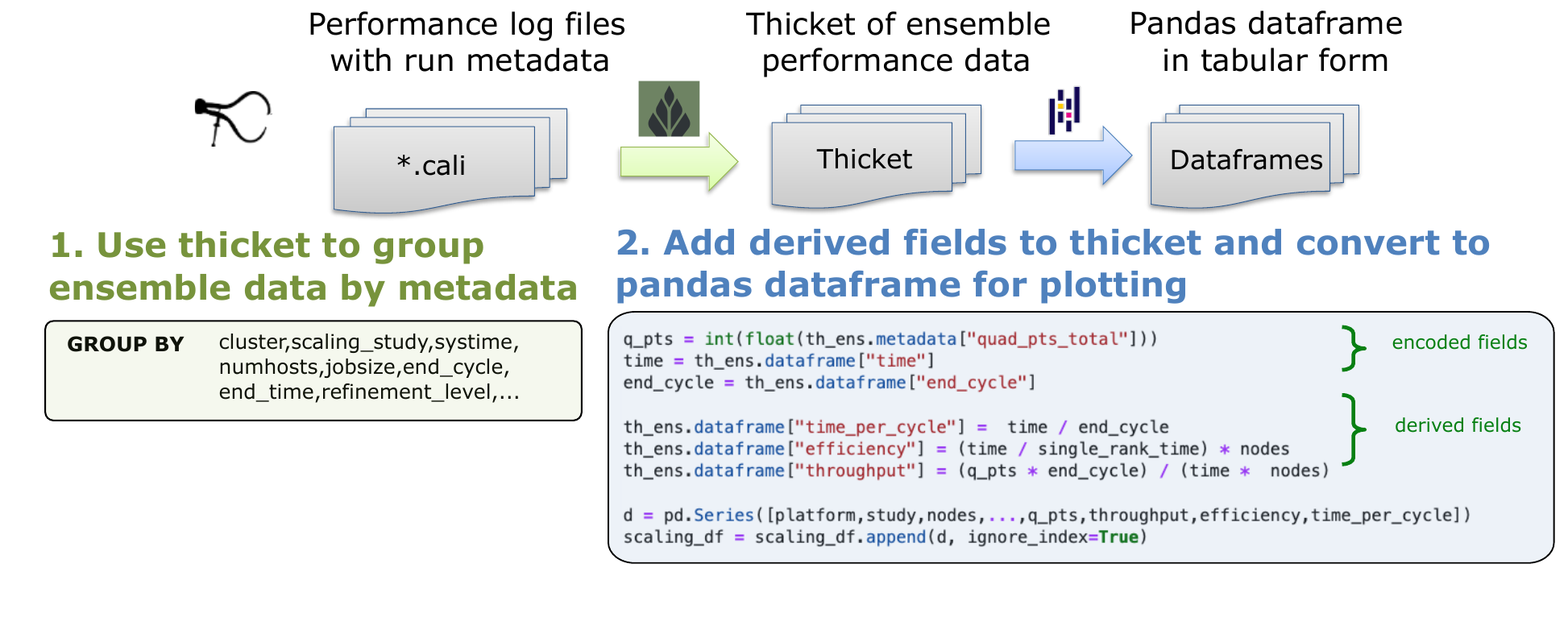}
        \caption{Our scaling study workflow postprocesses a collection of Caliper files 
                to extract the relevant runs and metrics for a given study.
                We use Thicket and pandas to convert the data into a simple tabular format and generate charts
                using matplotlib.}
        \label{fig:scaling_study_processing}
    \end{figure*}

    This section provides a brief high-level overview of our custom workflow 
    for setting up, running, and processing node-to-node scaling studies.
    While it is not the focus of this paper, it is worth mentioning the details that must manually
    be set up to run a scaling study and to filter the results into plottable charts.

    \begin{itemize}
        \item We used Maestro~\cite{maestro} to orchestrate the runs of our scaling study.
                We defined the relevant parameters in a yaml file and used jinja as an
                additional abstraction to fill in parameters related to each of the platforms
                and the various study parameters. See Figure~\ref{fig:scaling_study_jinja}.
        \item We used Caliper~\cite{boehme:sc2016} and Adiak~\cite{Axom-CS-infrastructure}
                to explicitly annotate sections of the code and metadata
                as described in~\cite{Boehme2021-isc}. This generates a Caliper file for each run, with
                associated metadata indicating the relevant quantities.
        \item We used Thicket~\cite{brink-2023-thicket} to read in and
              filter Caliper data based on metadata and create scaling charts for our data
              using matplotlib. Figure~\ref{fig:scaling_study_processing} illustrates the
              post-processing workflow we used for our studies.
    \end{itemize}

\section{Case Studies}
\label{sec:studies}

    This section discusses several case studies from the \marbl\ code in which
    we ran scaling studies to better understand code performance.
    We highlight insights from and features of the charts in this form.
    \marbl\ is a next-generation performance portable multiphysics simulation code developed at Lawrence Livermore National Laboratory
    for simulating high energy density physics (HEDP) and focused experiments driven by high-explosive, magnetic-
    or laser-based energy sources for applications including pulsed power and
    inertial confinement fusion (ICF)~\cite{Anderson2020}.
    \marbl\ is built on top of modular physics, computer science and math libraries.
    In particular, it makes extensive use of high-order finite elements from
    the MFEM library~\cite{mfem},
    has centralized computer science infrastructure provided by the Axom
    project~\cite{Axom-CS-infrastructure},
    uses Umpire~\cite{beckingsale2019umpire} as its memory abstraction later
    and RAJA~\cite{hornung2014raja,beckingsale2019raja} for its performance portability layer.
    These libraries enable \marbl's single-source codebase to perform well across a variety of
    CPU-based and GPU-based architectures.
    We ran studies on the following clusters:

    \begin{description}[leftmargin=1.5em, style=nextline, parsep=.5em]
        \item[Sierra (ATS-2)] LLNL's 125 petaflop ATS system. It has a Mellanox EDR InfiniBand interconnect and 4,320 compute nodes,
                each of which has two sockets containing 20-core POWER9
                processors, four NVIDIA Volta V100 16GB GPUs, and 256GB of memory.

        \item[Astra] A 2.3 petaflop system deployed under the Sandia Vanguard program.
                It has a Mellanox EDR InfiniBand interconnect and is composed of 2,592 compute nodes, of which 2,520 are user-accessible.
                Each node contains two sockets containing 28-core Cavium ThunderX2 64-bit Arm-v8 processors and 128GB of memory.

        \item[CTS-1] LLNL's CTS-1 clusters are commodity systems with Intel OmniPath interconnect,
                dual 18-core Intel Xeon E5-2695 2.1GHz processors and 128GB of memory per node.
                The system we tested on has 1,302 compute nodes and a peak of 3.2 petaflops.

        \item[CTS-2] LLNL's CTS-2 clusters are commodity systems with Cornelis Networks OmniPath interconnect,
                dual 56-core Intel Xeon Platinum 8480+ 2.0GHz processors and 256GB of memory per node.
                The system we tested on has 1,496 compute nodes and a peak of 10.7 petaflops.

        \item[EAS-3] LLNL's Early Access system for El Capitan. It has an HPE Slingshot 11 interconnect and 36 compute nodes,
                each of which has a single 64 core AMD Trento 2.0GHz processor,
                four AMD MI-250X 128GB GPUs, and 512GB of memory.

    \end{description}

    \begin{figure}[tbp]
        \centering
        \includegraphics[width=0.8\columnwidth]{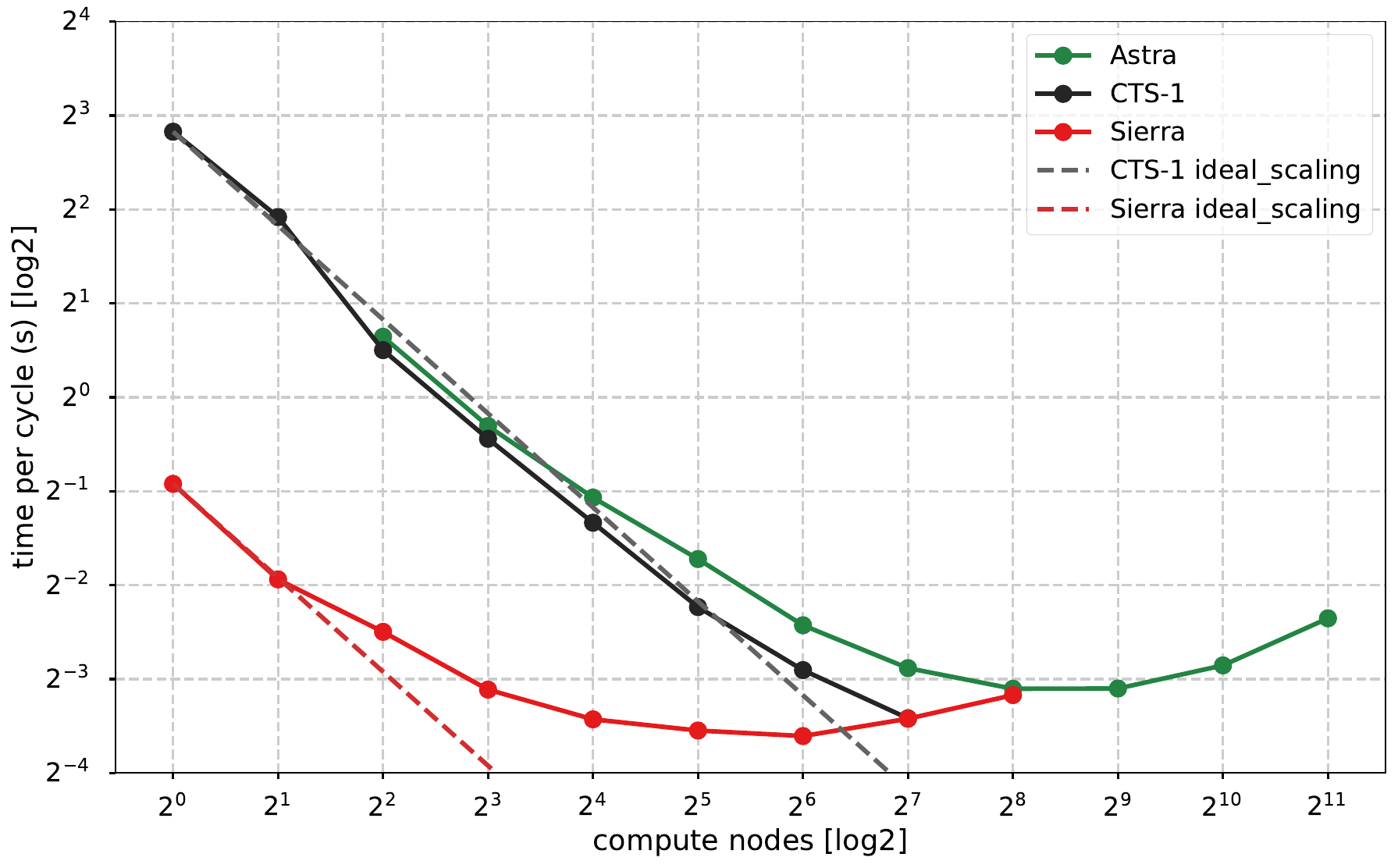}
        \caption{Node-to-node strong scaling for \marbl's Triple-Pt 3D problem with two levels of uniform refinement and TMOP.
            Each data point represents a 500 cycle Lagrange hydrodynamics simulation, with an ALE remap every 50 cycles.
            Data is plotted on a $log_2$--$log_2$ scale where the independent axis measures the number of compute nodes
            while the dependent axis measures the time per cycle (in seconds).
            \change{We also plot ideal strong scaling performance for the CTS-1 and Sierra series: a line with slope $-1$.
                 Data and figure from~\cite{Vargas2022-ijhpca} are reproduced here 
                 to highlight the layout and format of the chart.}
        }
        \label{fig:FY20-milestone-strong-scaling}
    \end{figure}

    \begin{figure}[tbp]
        \centering
        \includegraphics[width=0.8\columnwidth]{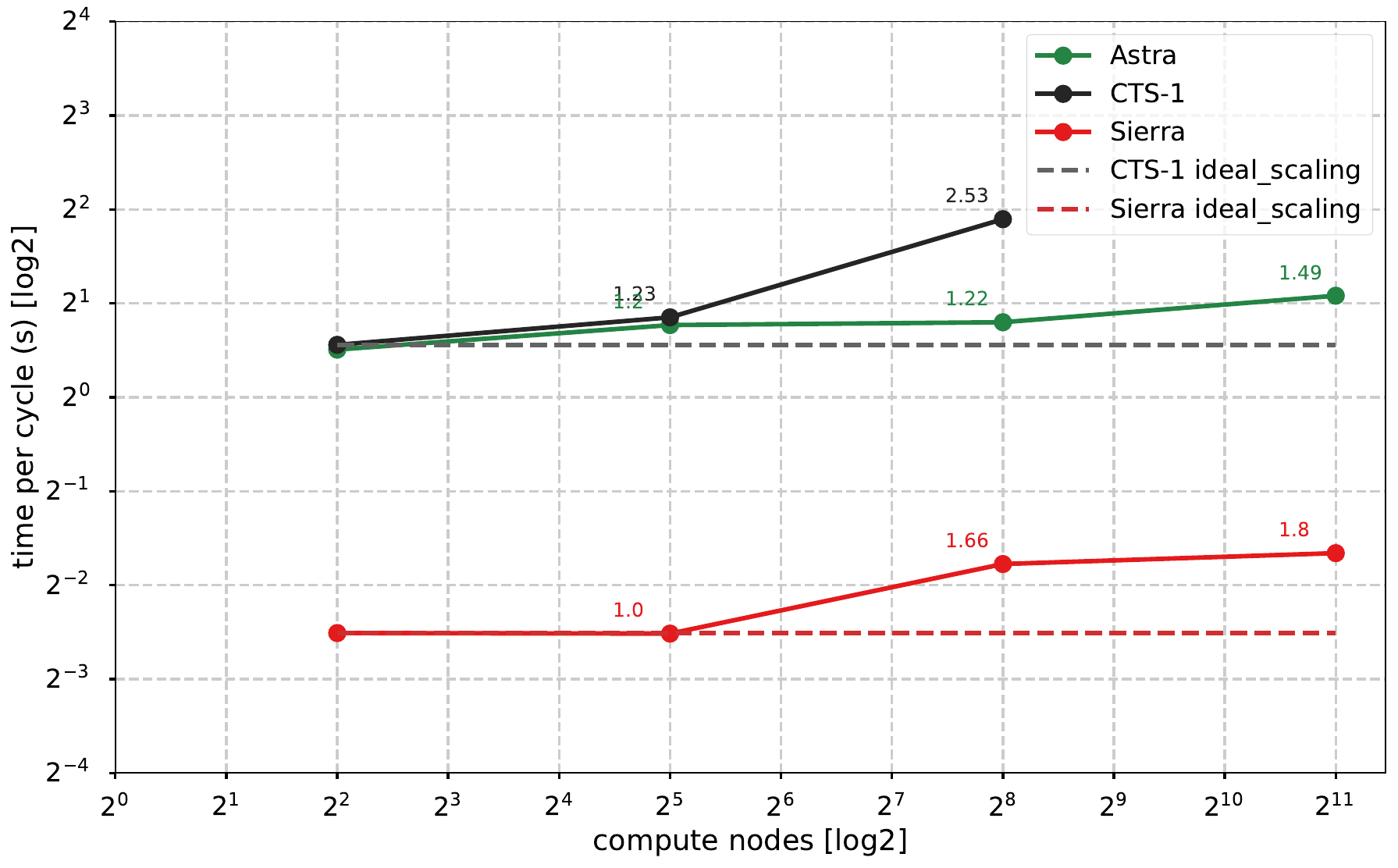}
        \caption{Node-to-node weak scaling for \marbl's Triple-Pt 3D problem on 4, 32, 256, and 2048 compute nodes.
            Each data point represents a 500 cycle Lagrange hydro simulation with an ALE remap step every 50 cycles.
            Data is plotted on a $log_2$--$log_2$ scale where the independent axis is the number of compute nodes
            and the dependent axis is the time per cycle (in seconds).
            The annotations indicate the inverse of the weak scaling efficiency with respect to the four node run on that platform.
            \change{We also plot ideal weak scaling, corresponding to an efficiency of 1.0, for CTS-1 and Sierra.
                 All data points in an ideal series would lie on a horizontal line.
                 Data and figure from~\cite{Vargas2022-ijhpca} are reproduced 
                 here to highlight the layout and format of the chart.}
        }
        \label{fig:FY20-milestone-weak-scaling}
    \end{figure}

\subsection{FY20 Project Milestone}
\label{sec:atdm_milestone}

    Our first case study was the impetus for the present work.
    For a project milestone in 2020, our team was tasked with designing a scaling study demonstrating
    strong and weak scaling of the \marbl\ codebase across significant portions of the GPU-based Sierra
    and the ARM-based Astra supercomputers, as well as on commodity CPU architectures.
    Our scaling studies used half of the Sierra cluster (2,048 nodes)
    as well as the full Astra cluster (2,496 nodes) and 256 nodes from CTS-1.

\begin{figure}[tbp]
    \centering
        \includegraphics[width=.85\columnwidth]{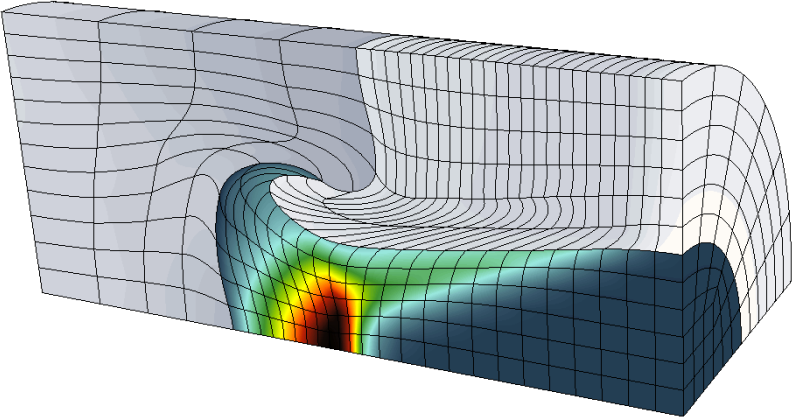}
        \caption{Visualization of \marbl's Triple-Pt 3D problem on a quadratic NURBS mesh}
        \label{fig:triple-pt-devilray}
    \end{figure}

    We ran this study on the Triple-Pt 3D hydrodynamics benchmark.
    For each platform, we ran 500 cycles of the Lagrange hydrodynamics simulation, with an ALE remap step every 50 cycles
    on an unstructured NURBS mesh (Figure~\ref{fig:triple-pt-devilray}).
    The mesh had 1,764 quadratic hexahedral elements at refinement level 0, each with 64 quadrature points,
    and successive factors of eight at each additional level of resolution,
    \eg 14,112 elements (with 903,168 quadrature points) at level 1
    and 112,896 elements (with 7,225,344 quadrature points) at level 2.
    Figure~\ref{fig:FY20-milestone-strong-scaling} and Figure~\ref{fig:FY20-milestone-weak-scaling}
    show node-to-node strong and weak scaling plots, respectively, of the Triple-Pt 3D problem
    with TMOP~\cite{dobrev2019target,knupp2012introducing} enabled for mesh optimization of \marbl's high-order mesh.
    While this scaling study was first presented in~\cite{Vargas2022-ijhpca},
    we reproduce it here to highlight several features of the charts:
    \begin{itemize}
        \item The use of a common independent variable --- the number of
                compute nodes --- allows for an easy comparison between
                three very different platforms. Using the number of processors/ranks/GPUs would make this much more difficult
                since each full node of Sierra is only using four MPI
                ranks, one rank per GPU.
                Similarly, for our Astra runs, we used two MPI ranks per node, with 28 OpenMP threads per rank.
        \item The $log_2$--$log_2$ axes combined with an extra
                \verb|ideal_scaling| line give the viewer an immediate sense of
                the expected performance. That is, a line with a slope of -1.
        \item Plotting the time per cycle rather than a more absolute time-to-solution factors out the arbitrary choice of 500 cycles in our setup.
                Thus, if we ran the simulation for more cycles, the plot would not change in any qualitatively meaningful way,
                and it helps the viewer get a sense of how long they might expect a simulation to run on a separate problem.
        \item Since each horizontal line corresponds to a factor of
                two, it is easy to see the speedup on the GPU-based
                Sierra compared to the CPU platforms (Astra and CTS-1) by
                counting the number of vertical boxes separating the lines. Each vertical box corresponds to a speedup by a factor of two.
        \item We also see nice initial strong scaling speedups on GPU platforms until there is not enough work to saturate
                the cores, after which the performance tapers off. 
                The initial $\sim\!15x$ speedup on a single node reduces
                to $\sim\!8x$ speedup at $2^3=8$ nodes, 
                and only about a factor of four at $2^5=32$ nodes after which the GPU performance continues to reduce.
                In practice, one would not want to strongly scale the problem on a GPU-based system to such a large extent.
        \item We see similar effects of strong scaling on Astra starting with around $2^7=128$ nodes.
    \end{itemize}

    Similar logic and considerations apply to our node-to-node weak scaling
plot in Figure~\ref{fig:FY20-milestone-weak-scaling}:
    \begin{itemize}
        \item The vertical separation highlights good estimates on the relative speedup/slowdown between platforms
                (by counting the vertical number of separating boxes).
        \item We add the (inverse of the) weak scaling efficiency as annotations on each data point to give the viewer a quick sense of
                how well each entry compares to its initial entry (in this
case, using four compute nodes).
        \item Using uniform refinement for our node-to-node weak scaling leads
to an easy setup. Each hexahedral element in the mesh
                is replaced by the eight elements obtained by bisecting the element along the x-, y- and z- axes. 
                However, this provides very few weak scaling datapoints; we
scale from four nodes to half of the Sierra cluster (2,048 nodes)
                using only three successive refinements.
        \item We observe that the code weak scales exceptionally well on Astra (with a slowdown of only 1.49 when running on 2,048 nodes)
                and we also see reasonable weak scaling up to half of Sierra (with a slowdown of 1.8 compared to the first datapoint).
    \end{itemize}

    \begin{figure}[tbp]
        \centering
        \includegraphics[width=0.8\columnwidth]{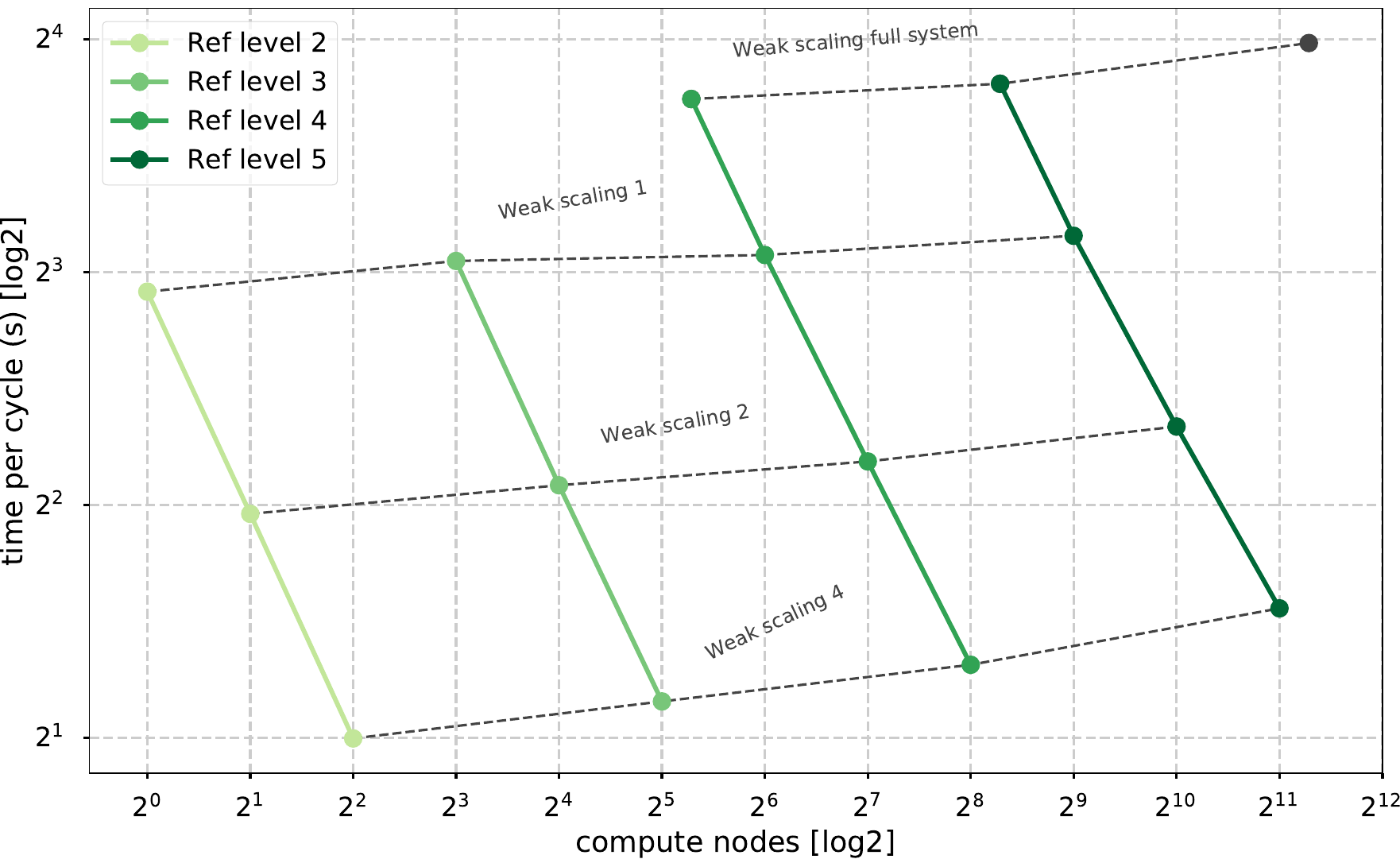}
        \includegraphics[width=0.8\columnwidth]{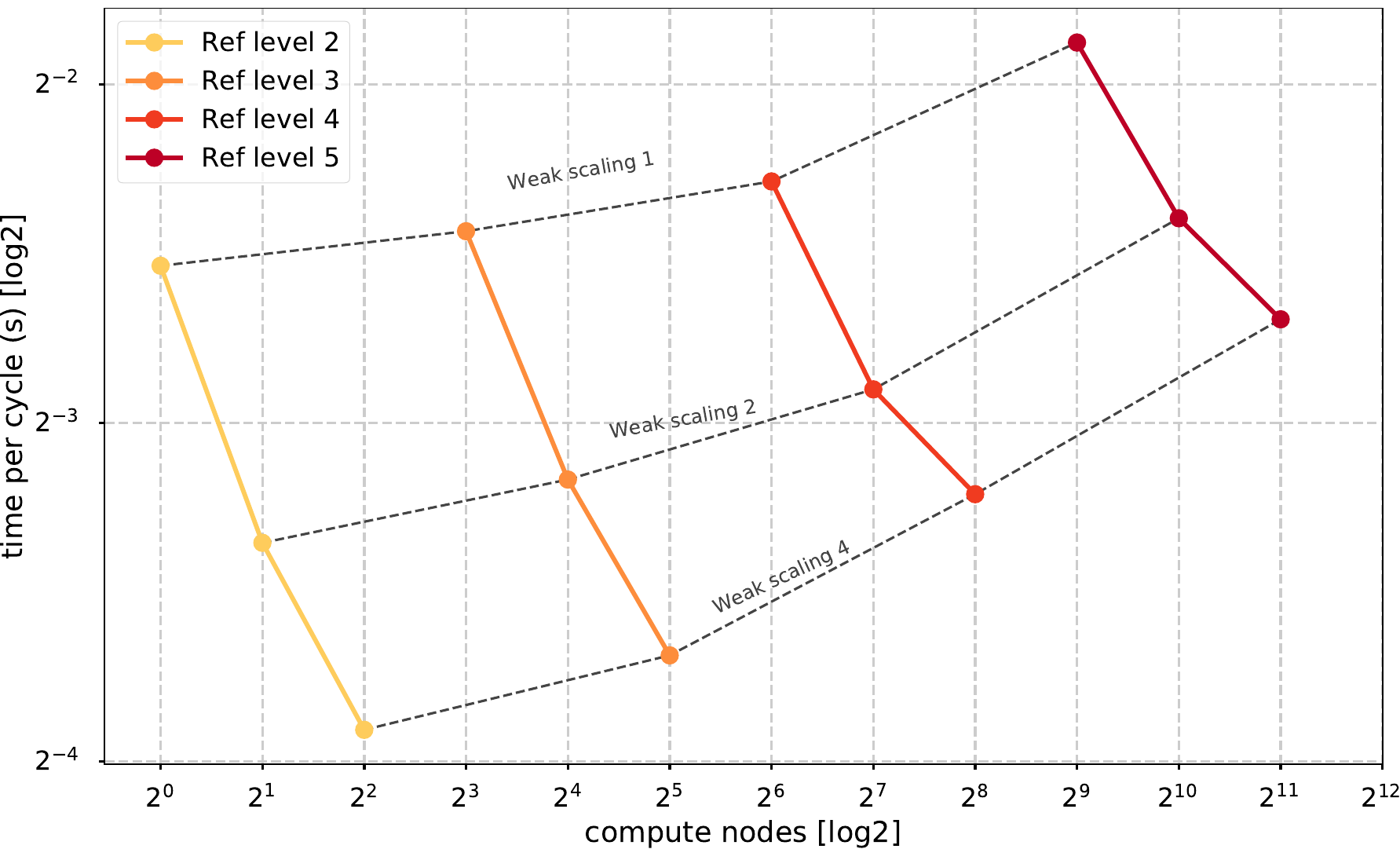}
        \caption{Strong-weak scaling plots for a scaling study on CPU-based Astra (top) and GPU-based Sierra (bottom) supercomputers.
                Data points connected by solid lines are from the same strong scaling series and have the same total degrees of freedom
                across a different number of ranks. Data points connected by dashed lines are from the same weak scaling series
                and have a fixed number of degrees of freedom on each compute node.
                A technical report provides more extensive details of this
study~\cite{Anderson2020}.
                }
        \label{fig:FY20-milestone-strong-weak-scaling}
    \end{figure}
    
    We also plot the node-to-node strong-weak scaling for a study without TMOP optimization in
    Figure~\ref{fig:FY20-milestone-strong-weak-scaling}
    for the ARM-based Astra cluster (top) and the NVIDIA-based Sierra cluster (bottom)
    This can provide users with a reasonable sense of how the code would be expected to perform on this problem
    if it were run on twice as many nodes (\ie by following the solid lines), or using roughly the same number of elements per node,
    by running a higher resolution version of the problem (\ie by following the dashed lines).
    More details about this study can be found in a technical report~\cite{Anderson2020}.

    We note that this study was run several years ago and the codebase has undergone significant
    improvements in the intervening years, including significant further GPU porting.

\subsection{Node-to-Node Throughput Study Comparing High-Order Runs on Different Systems}

    \begin{figure*}[tp]
        \centering
        \includegraphics[width=.8\textwidth]{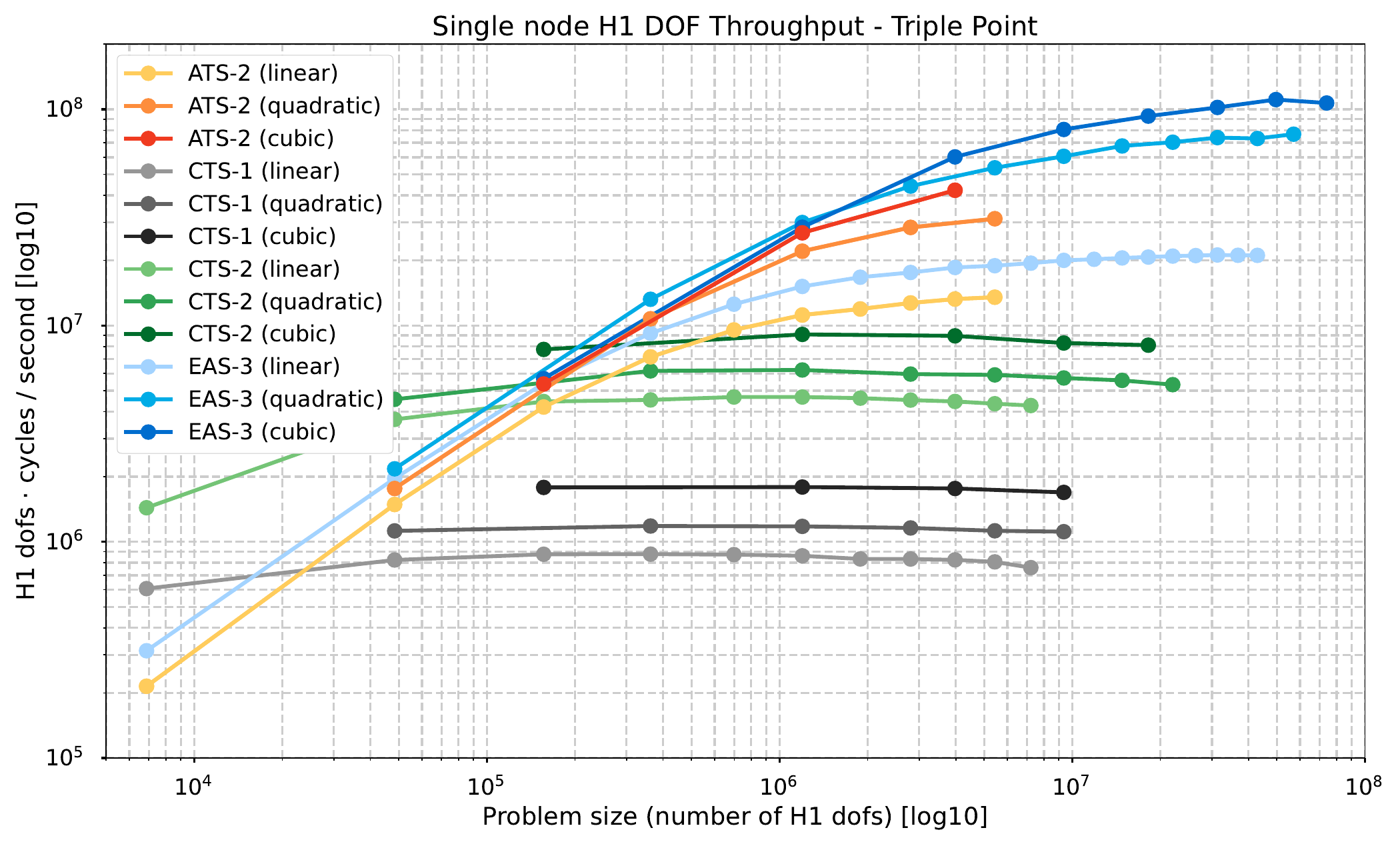}
        \caption{Comparing \marbl's Triple-Pt 3D node-to-node throughput scaling 
            on a single node of CTS-1, CTS-2, ATS-2, and EAS-3 using linear, quadratic, and cubic elements. 
            \change{Data and figure from~\cite{Stitt2024-jfe} are reproduced} here to highlight the layout and format of the chart.
        }
      \label{fig:throughput_123}
      \end{figure*}

      \begin{figure*}[tp]
        \captionsetup[subfigure]{justification=centering}
        \centering
        \subfloat[Matrix-based ALE remap]{
            \includegraphics[width=0.48\columnwidth]{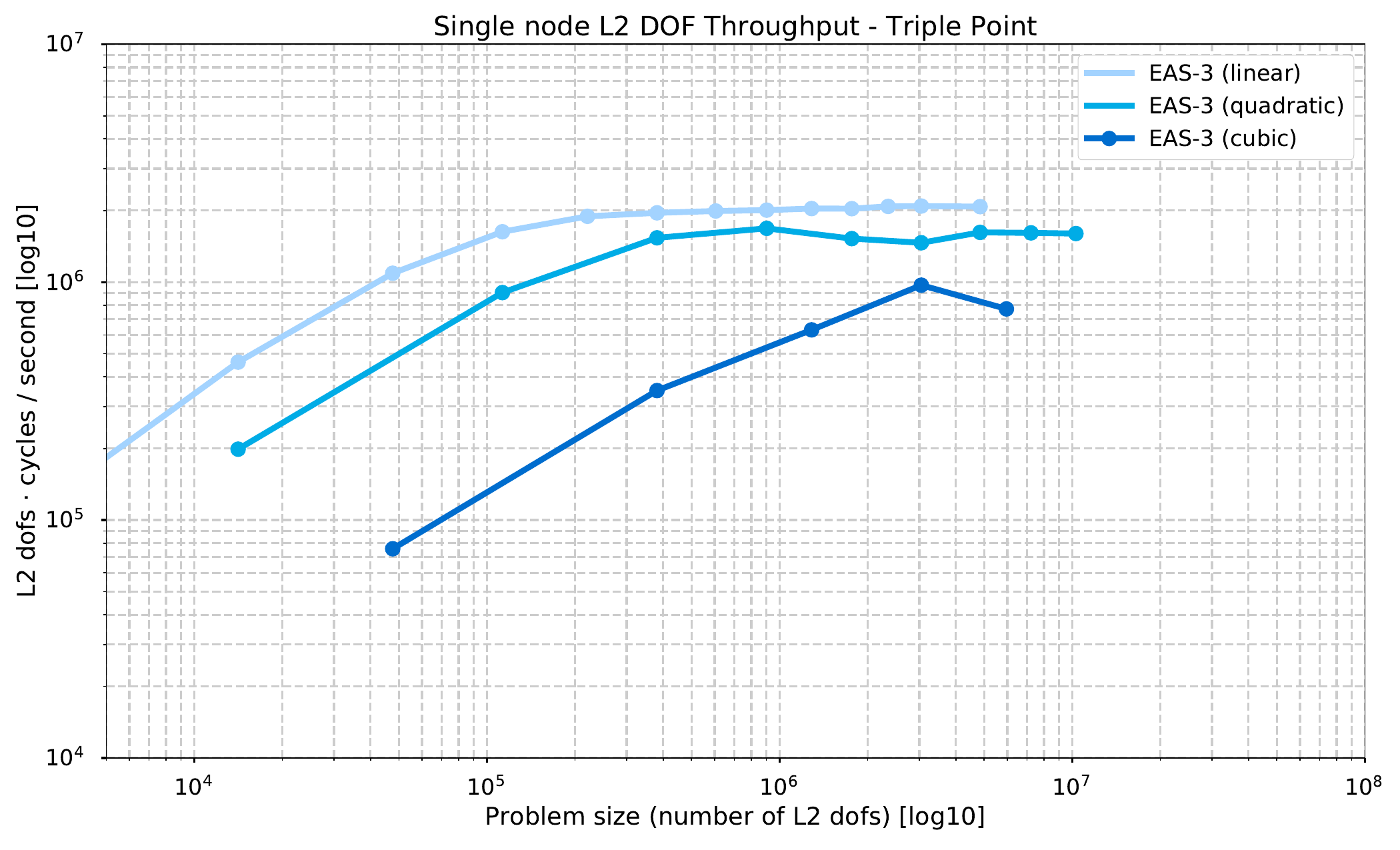}
            \label{fig:EAS3-throughput-matrix-based-remap}
        }
        \subfloat[Matrix-free ALE remap]{
            \includegraphics[width=0.48\columnwidth]{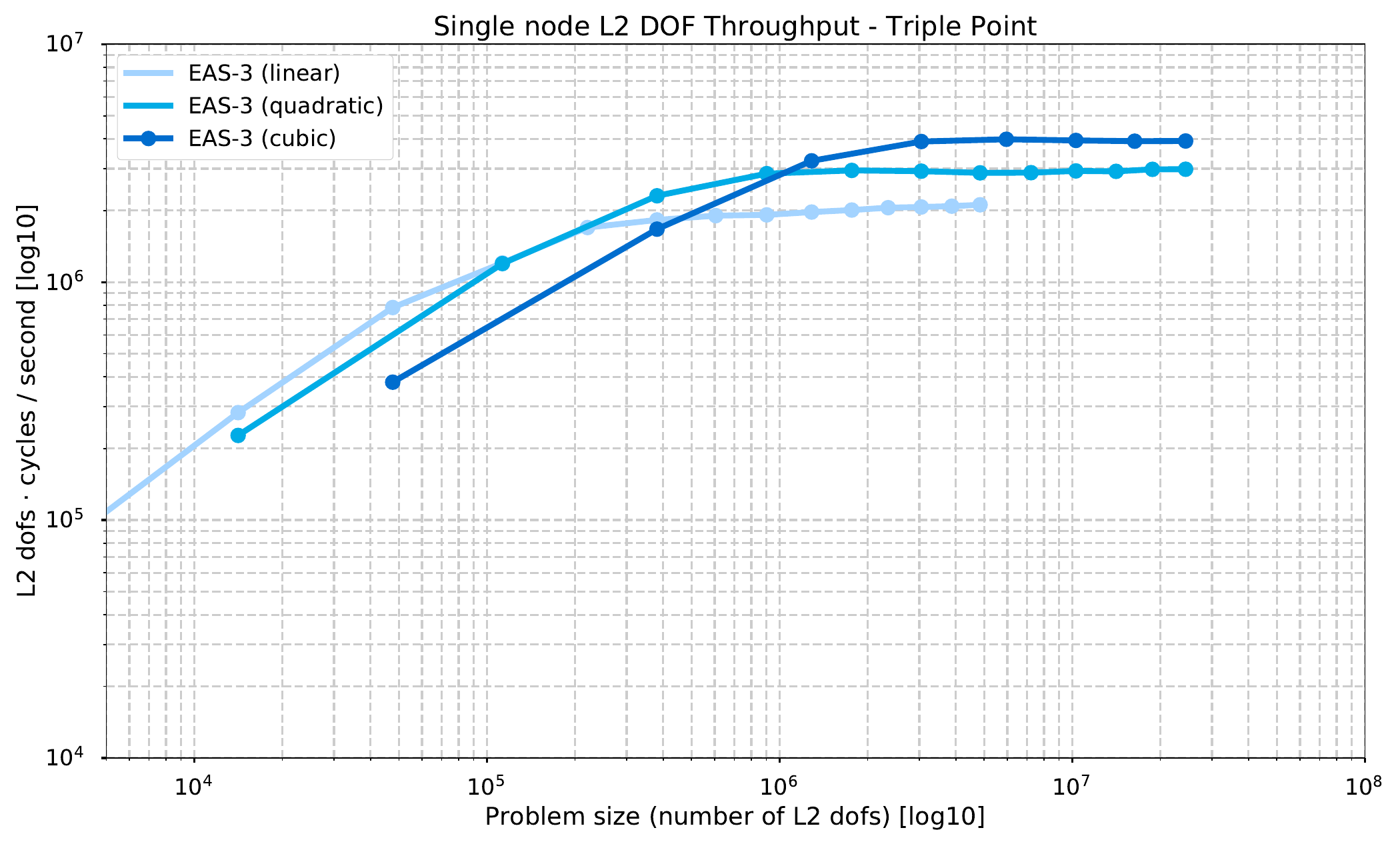}
            \label{fig:EAS3-throughput-matrix-free-remap}
        }
        \caption{Performance improvements achieved in \marbl's Triple-Pt 3D problem on EAS-3 (AMD MI250Xs) 
            by switching from (a) a matrix-based ALE remap formulation to (b) a matrix-free ALE remap formulation.
            \change{Data and figure from~\cite{Stitt2024-jfe} are reproduced} here to highlight the layout and format of the chart.
        }
        \label{fig:mf_remap_throughput_MI250x}
    \end{figure*}

    Our second case study uses node-to-node throughput scaling studies to compare \marbl's performance on the Triple-Pt 3D problem
    on several CPU-based and GPU-based systems. In addition to a better understanding of the performance of the codebase on different platforms,
    this study was especially concerned with understanding how the high-order code performs when running with different polynomial orders.
    Specifically, one of \marbl's distinguishing features is its use of high-order numerical discretizations, 
    which provide improved accuracy properties as well as scalability benefits from higher FLOP/byte ratios 
    -- each byte of data retrieved from memory has more floating point operations performed on it than a corresponding lower-order discretization.

    Our node-to-node throughput study (Figure~\ref{fig:throughput_123}) shows how the code performs 
    on different architectures and with different \change{polynomial} orders: linear (P1), quadratic (P2) and cubic (P3).
    We note that the data and charts were presented in~\cite{Stitt2024-jfe}, 
    and are reproduced here to highlight the features of the chart and its benefits in analyzing node-to-node throughput.

    As can be observed in Figure~\ref{fig:throughput_123}, on the CPU-based platforms (CTS-1 and CTS-2),
    \marbl\ saturates fairly quickly; the more powerful CTS-2 nodes achieve a higher throughput than their less powerful CTS-1 counterparts,
    but both throughput lines are relatively flat.
    The code achieves significantly higher throughput on the GPU-based platforms ATS-2 and EAS-3.
    The throughput scaling study also highlights the benefits of the increased
    memory available on EAS-3 nodes as compared to their ATS-2 counterparts.
    We can run problems that are an order of magnitude larger on EAS-3 nodes than on ATS-2 nodes before running out of memory.

    We can also see that the code achieves higher throughput on all platforms as the polynomial order is increased from linear to quadratic to cubic.
    \marbl's original matrix-based ALE Remap algorithm exhibited lower throughput for cubic meshes
    than for quadratic meshes due to its required full matrix assembly
operation~\cite{Vargas2022-ijhpca}. 
    \marbl's reformulation of the Remap algorithm using a matrix-free
algorithm~\cite{Stitt2024-jfe} 
    resolves this performance degradation, allowing the code to achieve higher throughput
    with increasing order. \change{Figure~\ref{fig:EAS3-throughput-matrix-based-remap} and
    Figure~\ref{fig:EAS3-throughput-matrix-free-remap} shows the throughput
    achieved before and after \marbl's reformulation of the Remap algorithm.}


\subsection{Cross-Platform Comparisons Using Different Library Features}

    \begin{figure}[tbp]
        \centering
        \includegraphics[width=0.8\columnwidth]{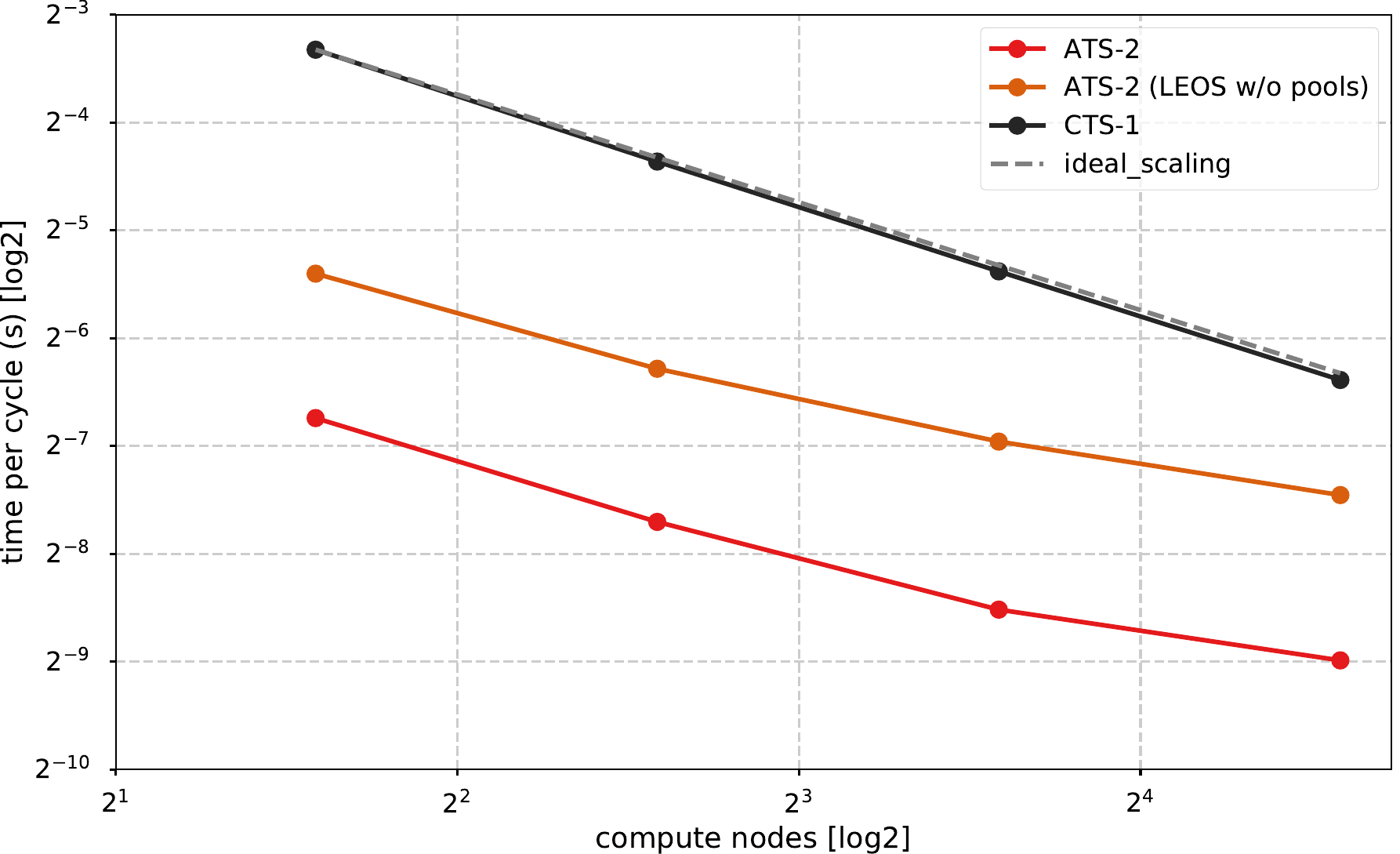}
        \caption{\marbl\ node-to-node strong scaling performance on CTS-1 vs. ATS-2 (Sierra) on the Shaped-Charge 3D problem.
          This study was concerned with the relative performance of the underlying equation-of-state library on GPUs
            when the latter was sharing pools memory with the main application.
            See~\cite{richards2021enhancements} for more details about the problem setup and comparisons.
        }
        \label{fig:PEM-study-shaped-charge-LEOS}
    \end{figure}

    Our third case study shows a cross-platform node-to-node strong scaling study of \marbl\ on the Shaped-Charge 3D problem 
    studying the performance of the code when integrated with a GPU-ported equation-of-state package, LEOS.
    
    A key aspect of this integration related to understanding the benefits of having 
    the host code, \marbl, and the equation-of-state library, LEOS, share preallocated memory pools
    compared to the simpler integration where each package used its own memory
    allocations on the device\change{.}
    This case study highlights how node-to-node scaling studies and the charts from Section~\ref{sec:n2n-scaling-studies}
    can help highlight the benefits of algorithmic changes within a code and/or the relative benefits of 
    running the code with different options.

    In particular, the node-to-node strong scaling chart in Figure~\ref{fig:PEM-study-shaped-charge-LEOS}
    compares the code's performance on a CPU-based platform (CTS-1), where memory pools were not necessary,
    against its performance on a GPU-based platform (ATS-2) when using shared memory pools (red line) 
    and when not using this feature (orange line).
    As is evident from the figure, we observed significant benefits (factors of 2x-4x improvements)
    across all scales when the libraries shared memory pools.
    More information about this study can be found in a technical report~\cite{richards2021enhancements}.

\subsection{\marbl\ vs. Containerized \marbl\ }

    \begin{figure}[tbp]
    \centering
        \includegraphics[width=0.8\columnwidth]{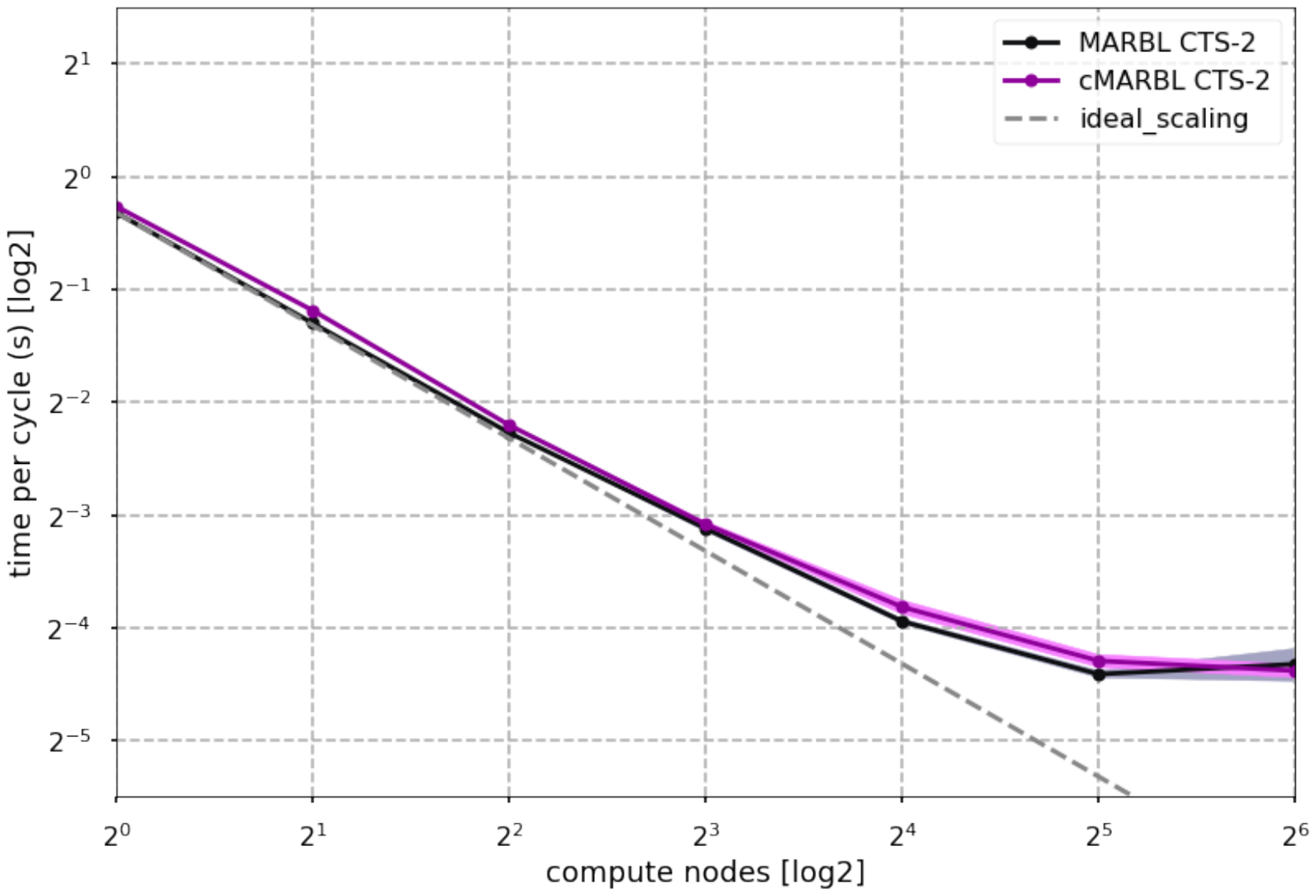}
        \caption{\marbl\ vs. c\marbl\ node-to-node strong scaling for the Triple-Pt 3D
            problem on the CTS-2 system. This study consisted of an ensemble of five \marbl\
            and five c\marbl\ runs. Each data point is the average of the five runs, and the
            shaded region surrounding each plot line represents the variability of the run.}
    \label{fig:marbl-vs-cmarbl}
    \end{figure}

    \begin{figure}[tbp]
    \centering
        \includegraphics[width=0.8\columnwidth]{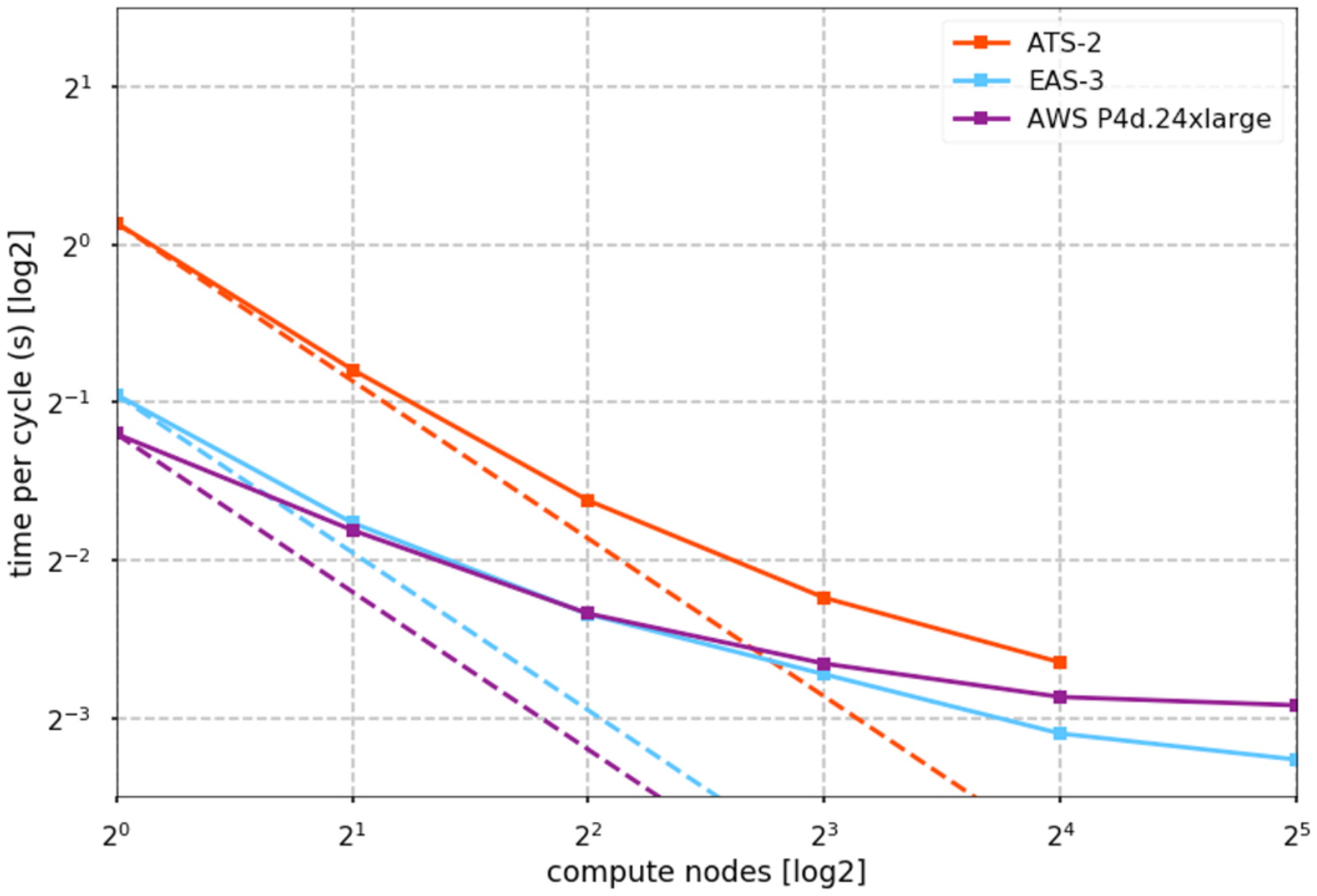}
        \caption{\marbl\ vs. c\marbl\ node-to-node strong scaling for a Shaped-Charge 3D problem.
            \marbl\ results are from the (on-prem) EAS-3 and ATS-2 systems. 
            c\marbl\ results are from an AWS P4d.24xlarge ParallelCluster instance.}
    \label{fig:aws-vs-eas3}
    \end{figure}

    As part of a project milestone in 2022, the team explored various containerization
    options for packaging the \marbl\ executable. Of particular importance was the performance of a containerized
    \marbl\ (c\marbl) compared to a native \marbl\ binary. Using various benchmark problems,
    we conducted a number of node-to-node strong and weak scaling studies of \marbl\ and c\marbl\ on both CPU
    and GPU architectures to confirm that there was negligible performance loss when running a
    containerized version of \marbl. Figure~\ref{fig:marbl-vs-cmarbl} shows the result of one such
    study using the Triple-Pt 3D problem on the CTS-2 system.

    The successful containerization of \marbl\ provided us with the opportunity to leverage cloud
    resources for various \marbl\ workloads. The ability to effectively compare the performance
    of cloud HPC resources to on-prem resources allows us to make better informed decisions when
    deciding which workloads we might burst to the cloud and the instance types those workloads
    would require. Figure~\ref{fig:aws-vs-eas3} shows the results of a node-to-node strong scaling study using
    a Shaped-Charge 3D problem running on our \change{GPU-based} on-prem EAS-3 system and similar GPU-based nodes
    on an AWS GovCloud ParallelCluster instance.

\section{Conclusion}
\label{sec:conclusion}

    \change{Cross-platform node-to-node studies are essential tools for understanding performance 
         in light of the increasing diversity of HPC architectures.}
    In addition to helping developers gauge the impact of their porting efforts,
    they factor into considerations about HPC acquisitions, especially when comparing early access systems where there
    are typically not enough nodes to run large scaling studies.
    They also help with comparisons when running simulations on cloud hardware where we must pay for each compute resource,
    and want to maximize performance per node rental costs.

    Despite this importance, there have been few resources in the literature about how to actually compare performance across platforms.
    This paper advocates for a single compute node as the unit for comparisons across systems
    and discusses how to set up such several types of node-to-node scaling studies.
    It also provides guidance on how to plot data from these runs to help analyze the results.

    However, much work remains. While we now have guidance on what the studies should contain, and how to set them up,
    actually setting these up and tracking data/metadata across runs requires a lot of manual work.
    Future work will focus on frameworks to more easily set up such studies.
    

    The diverse compute nodes on cloud computing clusters opens the door to interesting ``what-if'' questions
    not typically available to HPC center users.
    By comparing against the available cloud nodes, we would be able to easily get quantitative
    data in the form of node-to-node throughput charts on questions like:
    \begin{inparaitem}[]
        \item How would the code perform on a system with a different vendor's chips?
        \item How would the code perform on a node with twice as many flops, or with twice as much memory? Or with faster bandwidth memory?
    \end{inparaitem}

    We conclude by considering some differences between node-to-node scaling studies and traditional scaling studies.
    Whereas traditional scaling studies measure performance on the same platform as we parallelize a code,
    node-to-node scaling studies help with cross-platform performance comparisons for machines with different architectures.
    As such, we see their natural baseline unit measurement as an entire compute node, which effectively normalizes
    difference in node architectures and how users might run codes on them,
    such as comparing the four GPUs on the V100 nodes of Sierra
    to the scores of conventional cores on modern CPU-based architectures.
    We note that node-to-node scaling studies bake some intra-node communication costs
    into the initial single node measurements, so should be seen as an alternate way of measuring scalability,
    rather than a replacement for traditional scaling studies.

\section*{Acknowledgment}
  This work performed under the auspices of the U.S. Department of Energy by
Lawrence Livermore National Laboratory under Contract DE-AC52-07NA27344 and was supported by the LLNL-LDRD Program under Project No. 24-SI-005 (LLNL-JRNL-871833).
  We thank our colleagues Michael Glass, Bill Rider, Aimee Hungerford and David Daniel at Sandia National Laboratory and Los Alamos National Laboratory
  for discussions on setting up scaling studies for the shared project milestones discussed in Section~\ref{sec:atdm_milestone}.
  We thank Adam Kunen for discussions on throughput studies and sharing an early prototype for the node-to-node
  throughput chart presented in Section~\ref{sec:n2n-throughput-scaling-studies}.

\bibliographystyle{unsrt}
\bibliography{references}

\end{document}